# Modelo de Aprendizaje Biocibernético

# Biocybernetic Learning Model BLM

**Rommel Salas, MBA**
**Universidad del Turabo, Estudios Doctorales**
**Puerto Rico 2017**

**Abstracto**

La educación en el periodo digital en el que vivimos, está alcanzando retos nunca antes vistos, precedidos por fenómenos que involucran no solamente a unidades sociales tradicionales, sino también a las nuevas comunidades virtuales; innovar es difícil, es un reto, no obstante, hay que pensar en nuevos métodos de enseñanza que impacten a la actual generación de estudiantes, los mismos que llegan con nuevas necesidades y expectativas. La construcción del conocimiento desde el sujeto y el mundo virtual que lo rodea, establece la base para el desarrollo de un nuevo modelo de enseñanza, donde el salón de clase es la representación particular de un nuevo ecosistema físico-cibernético compuesto por las tres grandes dimensiones que forman parte de esta nueva convergencia tecno-social (humano – información – maquina); permitiendo una interrelación entre el estudiante, la información, la máquina y el profesor; usando métodos Biocibernéticos de influencia, control y replica, mediante el vector masivo de impacto ($i$); además, el desarrollo de nuevas estrategias asistidas por la cibernética y la actualización del contenido académico acorde al nuevo ambiente de enseñanza. De ahí la importancia de este estudio, el cual nos lleva a la necesidad de un nuevo modelo transformador de instrucción académica, el cual no se base en un conglomerado de herramientas tecnológicas, más establezca un nuevo modelo educativo y transformador, basado en el "pensamiento colaborativo" y la ubicuidad de la información, estableciendo así la relación entre el sujeto y objeto de estudio, permitiéndonos de esta manera establecer el nuevo paradigma educativo Biocibernético en el periodo digital.

Palabras claves: *Comunidades virtuales, Enseñanza, Teoría CTS, Academia, Periodo Digital.*

1. **Planteamiento del Problema**

Ya en el siglo XX los nuevos procesos productivos y tecnológicos basados en la globalización y modernidad (Barañano, 2010), establecen una brecha que se distancia de los principios hegemónicos de "Cultura individualista"; ya que esta es absorbida por la "globalización", entendiéndose, así como "Un fenómeno inevitable en la historia humana que ha acercado el mundo a través del intercambio de bienes y productos, información, conocimientos y cultura" (ONU, 2016).

Este proceso de globalización por medio de la gestión electrónica de información ha desarrollado dos polos: los que aun quieren mantener la identidad cultural enmarcada en tradiciones y estilos de vida establecido en el periodo tipográfico, y el otro que busca homogenización basada en la multiculturalidad como lo menciona (Gómez, 2001), así como el establecimiento de nuevos elementos en el proceso de interacción entre humanos, este punto de controversia ha ido subsanándose a través del desarrollo de las comunidades virtuales, las cuales establecen un proceso multicultural, y homogéneo, basado en una misma perspectiva, con vastos procesos de convivencia y coexistencia.

Con estos antecedentes, podemos establecer la existencia de un fenómeno que necesita ser observado para su respectiva investigación, el cual se centra en: a) Periodo digital en el que vivimos; b) Características generacionales del estudiantado; c) falta de criterio ambivalente de los órganos de control social hacia el establecimiento de normas Ciberéticas para establecer buenas prácticas hacia la convivencia entre el h+i+m (humano – información – maquina); e) El uso de métodos Biocibernéticos de influencia, control y replica; mediante el vector masivo de impacto (*i*) "información".

2. **Objetivo de la Investigación**

La construcción del conocimiento desde el sujeto y el mundo que lo rodea (Cantero, 2014), establece el desarrollo de una investigación que busca entender la interrelación entre el estudiante, la información, la máquina y el profesor; dentro del nuevo periodo digital, para lograr establecer un modelo que permita integrar al salón de clase en el ecosistema físico-cibernético, usando métodos Biocibernéticos de influencia, control y replica, mediante el vector masivo de impacto ($i$).

3. **Justificación**

En diferencia con otras áreas de investigación, en la cibernética social, el investigador es parte de la experiencia inmediata de una manera intrínseca, su relación permanente en los procesos de comunicación basados en tecnología, es enfocada claramente en la tricotomía de la CTS "h+i+m". Esto permite describir y explicar los significados vividos como menciona (Ruiz Olabuénaga, 1996), las experiencias diarias inmersas en la vida cotidiana, el predominio de opiniones sociales basadas en los "Hub" de colaboración y comunicación social, llamadas "Redes Sociales" y la frecuencia de ciertos comportamientos influenciados a base del vector masivo de impacto [i] "información" (Salas Guerra, 2016), nos lleva a tener una idea clara de lo que busca la fenomenología; la cual persigue mostrar los objetos tal como son, describirlos y no solo explicarlos (Universidad Alberto Hurtado, 1995).

La fenomenología en la cibernética social busca describir con mayor exactitud aquellos fenómenos en los cuales el humano es inmerso mediante el proceso convergente entre la tecnología y la social, buscando comprender los nuevos fenómenos sociales desde las perspectivas de las personas involucradas (Groenewald, 2004).

De la misma manera para (Protevi, 1994), las experiencias de la vida son individuales y únicas, las cuales parten de circunstancias generadas por los colectivos donde se desenvuelven; el periodo digital en el que vivimos, estos colectivos llamados "comunidades virtuales" establecen un nuevo ambiente de convivencia basada en un nuevo ecosistema físico-cibernético, ya que como menciona (McPhail, 1995) el sistema donde viven los individuos no es algo construido fuera de las experiencias vividas tanto a nivel individual como grupal.

4. **Limitaciones**

La principal limitación del estudio es el modelo tradicional de enseñanza basada en el periodo tipográfico, el cual tiene un arraigado fundamento en la memorización, adicionalmente a este factor, también nos encontramos con una falta de criterio gubernamental a establecer cambios profundos y completos en la infraestructura de enseñanza.

## 5. Marco Teórico

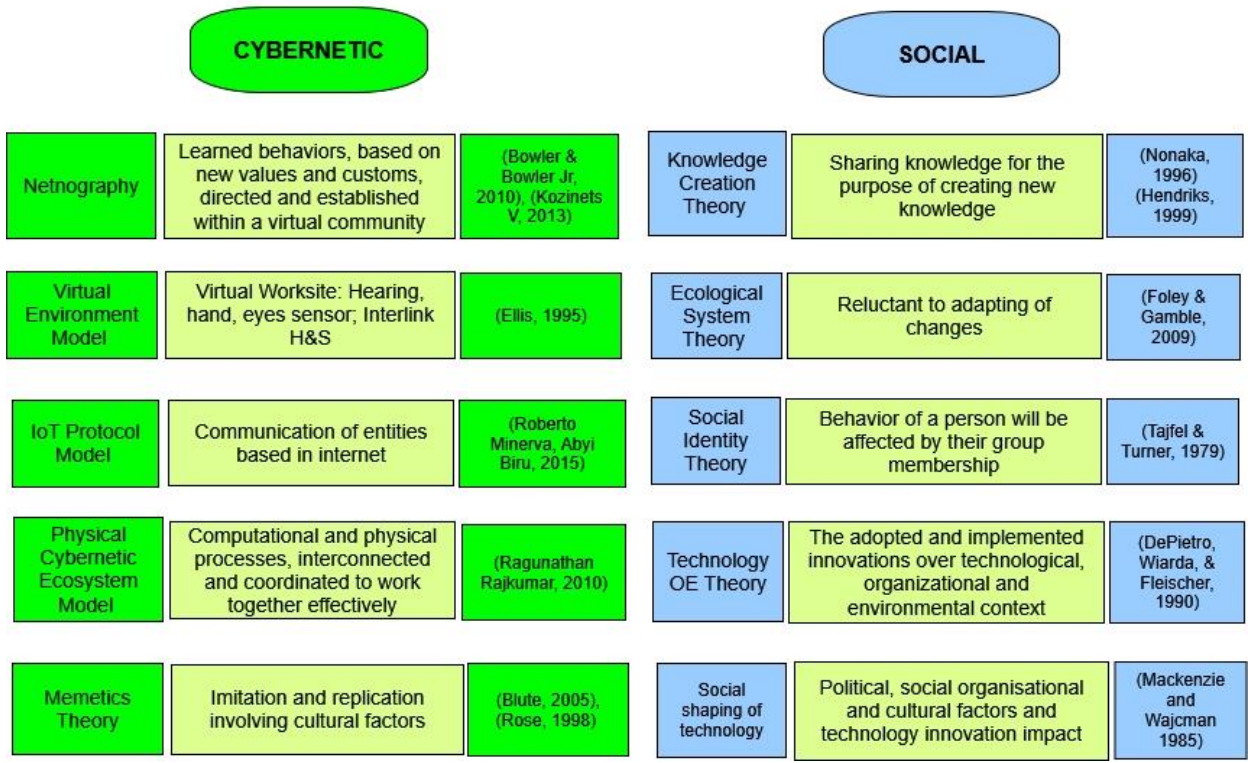

**5.1 La CTS como base epistemológica de la cibernética social**

El siglo XXI empieza con marcados cambios sociales, políticos y económicos a nivel mundial, pero lo más relevante es la "Convergencia Tecno – Social" CTS; el cual establece una nueva forma de vida humana basada en la convivencia (humano – información – maquina) dentro de un nuevo ecosistema físico – cibernético; estableciendo nuevas formas de comunicación humana que derivan en el Ciberantropo u hombre cibernético. Dentro del estudio de la convergencia tecno-social encontramos un enfoque de la comunicación social digital, que agilizo el desarrollo de un proceso de interacción (h + i + m); con el establecimiento de una manera nueva de compartir información dinámicamente.

La Teoría de la CTS, nos permite identificar, describir y establecer cuáles fueron los efectos del periodo digital en las unidades sociales, así como el establecimiento de un nuevo ecosistema de convivencia humano-maquina; conjuntamente con los futuros estudios de los niveles de impacto basados en; influencia, control y replica; mediante el vector masivo (i) "información", además la falta de criterio ambivalente de los órganos de control social hacia el establecimiento de normas Ciberéticas que establezcan el fundamento de la forma de vida del nuevo Ciberantropo.

Dentro de la CTS, disponemos de la variable (m) "maquina", la cual establece un enfoque de relación (informática – comunicación), el cual cubre objetos y estructuras en ambientes físicos, a lo cual actualmente se le denomina "Sistemas Físicos Cibernéticos" como lo menciona (Ragunathan Rajkumar, 2010), esto nos permite entender de una manera más objetiva los procesos de comunicación que coexisten en una comunidad virtual, donde se integra la relación (h + i + m). El camino es largo por recorrer, tenemos a la CTS que establece la ontología de la cibernética moderna, y sus respectivos enunciados epistemológicos, su aplicación es interdisciplinaria y multidisciplinaria.

**5.2 Base epistemológica educativa y la cibernética social**

Este estudio agrupa algunas teorías educativas que establecen la base conceptual del modelo de instrucción biocibernética, ya que como menciona Dewey "los estudiantes llegan al salón intensamente activos, el proceso de educación consiste en tomar esa actividad y orientarla" (Westbrook, 1993); esta cita refleja una gran verdad que vivimos hoy en día, nuestros estudiantes llegan a la Universidad con un sin número de expectativas que son experimentadas anticipadamente en sus comunidades virtuales.

### 5.2.1 La escuela activa de Jean Piaget

*Sólo la educación puede salvar nuestras sociedades de una posible disolución, violenta o gradual" (Piaget, 1934c, pág. 31)*

Piaget fue uno de los defensores del establecimiento de un nuevo modelo de educación, el cual buscaba un desarrollo que innove la importancia del principio de libertad, actividad e interés del estudiante (Piaget, 1999). El enfoque de Piaget establece de acuerdo a (Villar, 2003), un resultado en el origen del conocimiento basado en la interacción entre el sujeto y objeto, el gradual ajuste entre el sujeto y el mundo exterior, el cual hoy en día no solamente abarca aspectos físicos, sino también nuevas formas de comunicación humana basadas en la interrelación humano-información-maquina (Salas Guerra, 2016), lo cual nos lleva a entender que esta aportación epistemológica de Piaget nutre esta relación holística.

Esta construcción de conocimientos denota un proceso biológico de asimilación basado en la integración de elementos disponibles en la estructura de un nuevo ambiente de enseñanza Biocibernético, donde la interacción entre el sujeto y su entorno establecen el proceso de adquirir, refinar conocimientos y destrezas (Arocho, 2017). Lográndose de esta manera desarrollar nuevas experiencias educativas por medio del involucramiento de todos los componentes del nuevo modelo de enseñanza Biocibernético compuestos por el estudiante-información-maquina-profesor.

### 5.2 2 La teoría del conocer de John Dewey

Una de las afirmaciones más enriquecedoras de Dewey establece la derivación exitosa de la práctica para la obtención del conocimiento la misma que se empleara para perfeccionar métodos que tienen menos éxito (Cadrecha, 1990),

### 5.2.3 El modelo instrucción centrada en el alumno o LCI

Este modelo describe los esfuerzos para el desarrollo de métodos de enseñanza que se basan en la transmisión de Conocimiento (Paris & Combs, 2006), la cual está basado en la investigación. Este enfoque establece un énfasis en prácticas efectivas de enseñanza, las cuales consisten en establecer al estudiante como punto de partida para el diseño curricular; los profesores y los estudiantes son coparticipes en el proceso de aprendizaje; además el maestro promueve un intenso compromiso en el plan de estudios, permitiendo al estudiante desarrollar habilidades de resolución de problemas, beneficiando al estudiante con adquisición, retención y transferencia de conocimiento, autoconciencia, pensamiento crítico, y la motivación (Malone, 2008).

6. **Framework**

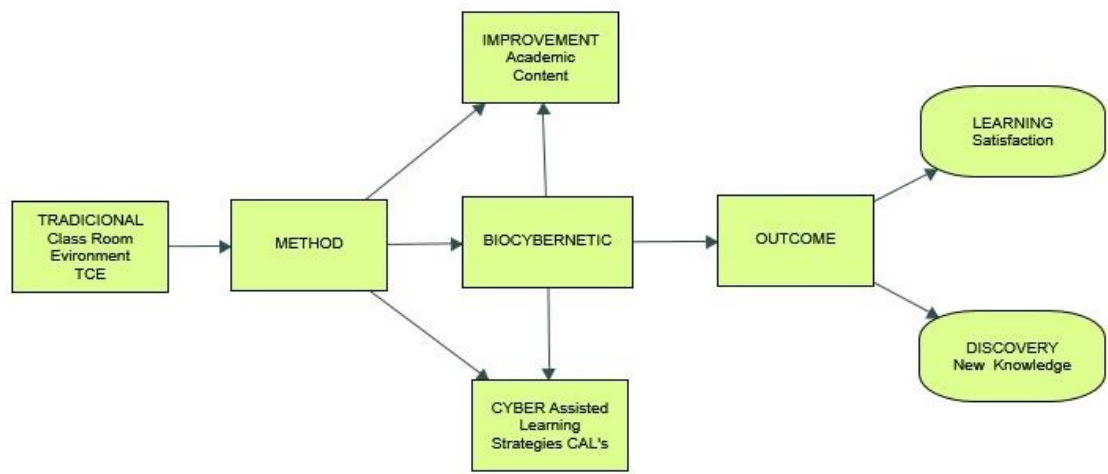

7. **El Método Cualitativo como base del estudio Biocibernético**

Este estudio con base interpretativo – explicativo se fundamenta en dos modalidades de la investigación cualitativa, las cuales envuelven las teorías de la cultura y los enfoques filosófico-metodológicos (Sandoval Casilimas, 1996); dentro de las cuales el estudio combina la: Teoría fundamentada, etnometodología, el interaccionismo simbólico, la netnografia (etnografía digital), hermenéutica y la fenomenología. En relación al levantamiento de data, el estudio usa data semi-estructurada, que parte de una pregunta de investigación.

## 7.1 Base epistemológica cualitativa para el estudio del Modelo Instruccional Biocibernético

Para (Universidad Alberto Hurtado, 1995), la investigación estratégica cualitivista, en diferencia a la cuántica cuantitavista, tiene un proceso secuencial y distributivo basado en la adaptación de las características particulares de aquello que se va a estudiar. Para (Quintana, 2006) este método investigativo se centra en la comprensión de una realidad vista desde los aspectos particulares como fruto de un proceso social; para (Ruiz Olabuénaga, 1996) estudia la realidad en un contexto natural, tal y como sucede, esto con un enfoque en los significados que tiene para las personas implicadas en el fenómeno a estudiar; y para (López-Herrera & Salas-Harms, 2009) se caracteriza por evitar la cuantificación para basarse en descripciones narrativas, estudiando los contextos estructurales y situacionales. Por lo tanto establecer al método cualitativo como la base de este estudio nos permite trabajar con objetos visibles y experienciables (Padrón, 2007)

## 7.2 La Teoría Fundadamentada

Para (Glaser & Strauss, 1967), el tema básico de estudio es el descubrimiento de la teoría datos obtenidos sistemáticamente con una base rigurosa dentro de la investigación social, ya que el investigador luego del proceso de observación (Descriptiva, focalizada y selectiva), inicia uno de los trabajos más arduos que es la organización de información obtenida dentro de las técnicas etnográficas, para (Alicia et al., 2011) esta teoría permite establecer una base documentológica basada en la observación e interpretación de la realidad observada; la cual de acuerdo a (Quintana, 2006) es obtenido a través del dialogo y contacto directo con el investigado.

## 7.3 El interaccionismo Simbólico

El cual como menciona (Hernández Carrera, 2014) es aquel que se interesa en todos los significados subjetivos mediante los cuales los individuos de estudio dan a los objetos; el interaccionismo simbólico comprende que son las personas las que actúan en la sociedad y no la sociedad en las personas (Carmen, 2006); estableciendo que de los contextos de interacción se derivan los significados en las personas, superponiéndose la comprensión de los significados en una explicación causal como lo establece (Cantero, 2014).

**7.4 La Etnometodología**

Establece su base en las cuestiones cotidianas y la manera como se generan (Hernández Carrera, 2014), buscando de esta forma entender cómo se establecen los procesos de interacción social entre las personas, dándole sentido y significado a las prácticas habituales en la sociedad, estableciéndose que las acciones de las personas se pueden explicar únicamente en referencia al contexto en el cual se desarrollaron (Ruiz Olabuénaga, 1996); bajo esta perspectiva metodológica el objetivo es describir y comprender los conceptos obtenidos en la interacción social atreves del análisis de la data significativa obtenida en el estudio, (Garfinkel, 1967) establece como objetivos de estudio en la etnometodología las actividades prácticas, las circunstancias prácticas, y el razonamiento sociológico practico.

**7.5 La Fenomenología**

Para (Sandoval Casilimas, 1996), este pilar conceptual se fundamenta en el enfoque sobre experiencias vividas, establecidas por el análisis de los datos obtenidos en fuente no estructuradas, estructuradas y semi-estructuradas, con el objetivo de establecer procesos descriptivos y legítimos de los datos obtenidos (Las, 2012); este enunciado yace en las experiencias personales del sujeto de estudio (Padrón, 2007), además como menciona (Sandoval Casilimas, 1996) permite entender la lógica de los sujetos de estudio en interacción con sus condiciones de vida.

**7.6 La Netnografia o Etnografía Digital**

El individuo al ser miembro de una unidad social, es objeto de estudio de la "etnografía", la cual busca establecer un proceso de "observación" directa (Murillo, 2016), que permita el levantamiento de información para su posterior análisis. Como menciona (Salas Guerra, 2016) el desarrollo investigativo en el área etnográfica, en los últimos años fue impactada por la tecnología, el cual es estudiado como un fenómeno que transformó la sociedad, de ahí la importancia de la etnografía en este estudio; el cual proviene del griego "ethnos" (Tribu o Pueblo), y "grapho" (Yo escribo), el cual busca describir el modo de vida de un grupo de individuos; así como la descripción del modo de vida de una raza o grupo de individuos. Para (Thomas & Blattberg, 2015) la etnografía se basa en el conocimiento de lo local, lo particular y especifico buscando con frecuencia la obtención de conocimiento fundamentado.

Dentro de todo proceso de investigación científica; las herramientas y modelos que se usan son de suma importancia; la investigación de campo como lo menciona (Bartis, 2004), "requiere la observación de primera mano, mediante documentación de lo que el investigador observa y escucha en un sitio particular"; por esta razón el modelo de investigación que usa este estudio es el Netnografíco, para entender el mismo es importante establecer procesos jerárquicos sociales que parten de un macro concepto de la antropología social que establece la distinción entre identidad individual y la identificación colectiva, la cual de acuerdo a (Barañano, 2010), establece que los "significados culturales, son sobre todo individuales e interculturales"; esta premisa nos lleva a dilucidar atributos que establecen una de las bases de este estudio.

El método de investigación científica basada en Netnografía, nos permite realizar estudios enfocados en: conductas aprendidas, estos comportamientos basados en nuevos valores y costumbres, dirigidos y establecidos dentro de una comunidad virtual, de acuerdo a (Bowler, 2010). Esta interacción vinculante (humano – información - maquina) o (h + i + m), nos permite entender factores ontológicos de la sociedad y la cibercultura; aunque el énfasis de muchos investigadores es establecer variantes y características entre una comunidad virtual y una tradicional; debemos ir más allá de los antiguos paradigmas del paralelismo entre la tecnología y la sociedad, estableciendo de esta manera un criterio unificado en cuanto a una nueva forma de vida tecno incluyente, que es el núcleo del estudio de la CTS.

**7.6 La Hermenéutica**

Etimológicamente, como menciona (Ghasemi, Taghinejad, Kabiri, & Imani, 2011), la palabra "Hermenéutica" proviene de la figura mitológica griega "Hermes" quien fue el responsable de interpretar los mensajes de Zeus para los otros dioses y diosas.

Este método interpretativo (Martínez, 2006) busca desarrollar un proceso de observación para de esta manera encontrar un significado; dentro de esta misma perspectiva (Sandoval Casilimas, 1996) manifiesta que el objetivo de esta es desarrollar la observación de estilos de vida y condiciones sobre la perspectiva del presente y pasado de culturas, grupos o individuos.

Por medio de la Hermenéutica se hace explicita la estructura de una "una situación" (Zalta, 2009), siendo esta la base del método interpretativo, permitiéndonos de esta manera entender que la esencia de información que se observa en un "segmento" o pieza de información, no necesariamente es la "representacional" de lo que vemos, leemos o escuchamos, en nuestro contacto cibernético. Para (Lewis & Grimes, 1999) las técnicas Hermenéuticas nos ayudaran a identificar los significados compartidos por los miembros sub-culturales.

Este proceso nos permitirá identificar de acuerdo a (Thompson, 1997) significados culturales personalizados que constituyen su sentido de identidad propia, y el significado de acontecimientos y experiencias específicas de la vida.

## 8. Conclusiones e Implicaciones

Este trabajo cualitativo experimental se inicia con una revisión literaria teórica-conceptual estableciendo la base epistemológica para la aplicación de la metodología netnográfica en el proceso de recolección y análisis de data no estructurada en comunidades virtuales donde los estudiantes están inmersos, los instrumentos de operabilización que se usarán para adquirir la misma tendrán intervención directa del participante.

**Referencias**

Alicia, D., Preparado, I., Hernández, :, Herrera, J. G., Martínez, L., Páez, R., … Auxiliadora, M. (2011). Teoría fundamentada, 1–39.

Arocho, W. C. R. (Universidad D. P. R. (2017). Materiales en línea. Proyecto para el Desarrollo de Destrezas de Pensamiento. Actualidad de las ideas pedagógicas de de Jean Piaget y Lev Vygotsky: Invitación a la lectura de los textos originales, 1–8. Retrieved from www.pddpupr.org

Cadrecha, M. A. (1990). John Dewey: propuesta de un modelo educativo. *Aula Abierta*.

Cantero, D. S. M. (2014). Teoría fundamentada y atlas.ti: Recursos metodológicos para la investigación educativa. *Revista Electronica de Investigacion Educativa*, *16*(1), 104–122.

Carmen, D. la C. B. (2006). La teoría fundamentada como herramienta de análisis. *Cultura de Los Ciudadanos*, *2 Semestre*(20), 136–140.

Garfinkel, H. (1967). Studies in ETHNOMETHOOOLOGY, 288. http://doi.org/10.4324/9781315775357

Ghasemi, A., Taghinejad, M., Kabiri, A., & Imani, M. (2011). Ricoeur's Theory of Interpretation : A Method for Understanding Text ( Course Text ), *15*(11), 1623–1629.

Glaser, B. G., & Strauss, A. L. (1967). *The Discovery of Grounded Theory: Strategies for Qualitative Research. Observations* (Vol. 1). http://doi.org/10.2307/2575405

Groenewald, T. (2004). A Phenomenological Research Design Illustrated. *International Journal of Qualitative Methods*, *3*(1), 1–26. http://doi.org/Retrieved from: http://www.ualberta.ca/~iiqm/backissues/3_1/html/groenewald.html

Hernández Carrera, R. M. (2014). La Investigación Cualitativa a Tr Avés De Entrevistas: Su Análisis Mediante La Teoría Fundamentada. *Cuestiones Pedagógicas*, *23*, 187–210.

Las, X. (2012). Ontología Y Epistemología En La Investigación.

Lewis, M. W., & Grimes, A. I. (1999). METATRIANGULATION: BUILDING THEORY FROM MULTIPLE PARADIGMS. *Management Review*, *24*(4), 672690.

Malone, D. M. (2008). Inquiry-based early childhood teacher preparation: The personal learning plan method. *Early Childhood Education Journal*, *35*(6), 531–542. http://doi.org/10.1007/s10643-008-0237-4

Martínez, M. (2006). La investigación cualitativa: síntesis conceptual. *Iipsi*, *9*(1), 123–146. http://doi.org/1560 - 909X

McPhail, J. C. (1995). Phenomenology as philosophy and method. *Remedial and Special Education*, *16*(3), 159–165.

Padrón, J. (2007). Tendencias epistemológicas de la investigación científica en el Siglo XXI. *Cinta de Moebio. Revi Epistemol*, (28), 1–28. Retrieved from http://www.derechoinformatico.uchile.cl/index.php/CDM/article/viewArticle/25930

Paris, C., & Combs, B. (2006). Lived meanings: what teachers mean when they say they are learner-centered. *Teachers and Teaching*, *12*(5), 571–592. http://doi.org/10.1080/13540600600832296

Piaget, J. (1999). *De la Pedagogia*.

Protevi, J. (1994). Notes on Antonio Damasio's Descartes' Error: Emotion, Reason and the Human Brain.

Quintana, A. (2006). Metodología de Investigación Científica Cualitativa. *Psicologia: Tópicos de Actualidad*, 47–84.

Ruiz Olabuénaga, J. I. (1996). Metodología de la investigación cualitativa.

Salas Guerra, R. (2016). Antropologia de la Informática Social: Teoria de la convergencia Tecno-Social. *VirtualEduca Conference*.

Sandoval Casilimas, C. (1996). *Especialización en teoría, métodos y técnicas de investigación social. Módulo.* http://doi.org/958-9329-18-7

Thomas, S., & Blattberg, R. C. (2015). Journal of Marketing Research, *41*(1), 31–45.

Thompson, C. J. (1997). Interpreting consumers: A hermeneutical framework for deriving marketing insights from the texts of consumers' consumption stories. *Journal of Marketing Research*, *34*(4), 438–455. http://doi.org/10.2307/3151963

Universidad Alberto Hurtado. (1995). El Proceso de Investigación Cualitativa, 40.

Villar, F. (2003). El enfoque constructivista de Piaget (Capítulo 5). *Proyecto Docente. Psicología Evolutiva Y Psicologia de La Educación*, 262–305. Retrieved from http://www.ub.edu/dppsed/fvillar/principal/pdf/proyecto/cap_05_piaget.pdf

Westbrook, R. (1993). John Dewey. *Perspectivas: Revista Trimestral de Educación*

*Comparada, XXIII*, 289–305.

Zalta, E. N. (2009). Hans-Georg Gadamer Stanford Encyclopedia of Philosophy. *Stanford Encyclopedia of Philosophy*, 588–599.